\documentclass[12pt,a4paper,notitlepage]{article}
\usepackage[a4paper, total={6.5in, 9in}]{geometry}

\usepackage{biblatex} 
\usepackage[normal]{caption}
\usepackage{subcaption}

\usepackage{booktabs}
\usepackage{bm}
\usepackage{float}
\usepackage{longtable}
\usepackage{enumitem}
\setlist[description]{leftmargin=\parindent,labelindent=\parindent}

\newcommand{\textbiglparen}{$\bigl($}
\newcommand{\textbigrparen}{$\bigr)$}

\usepackage{graphicx}
\usepackage{wrapfig}

\usepackage[dvipsnames]{xcolor}
\usepackage{xargs} 
\usepackage[textwidth=2.8cm,colorinlistoftodos,prependcaption,textsize=footnotesize,disable]{todonotes}

\newcommandx{\dmitry}[2][1=]{\todo[linecolor=red,backgroundcolor=red!25,bordercolor=red,author=Dmitry,#1]{ #2 }}
\newcommandx{\sergio}[2][1=]{\todo[linecolor=blue,backgroundcolor=blue!25,bordercolor=blue, author=Sergio,#1]{ #2 }}
\newcommandx{\dario}[2][1=]{\todo[linecolor=green,backgroundcolor=green!25,bordercolor=green,author=Dario,#1]{ #2 }}
\newcommandx{\tommaso}[2][1=]{\todo[linecolor=violet,backgroundcolor=violet!25,bordercolor=violet,author=Tommaso,#1]{ #2 }}
\newcommandx{\jam}[2][1=]{\todo[linecolor=yellow,backgroundcolor=yellow!25,bordercolor=violet,author=jam,#1]{ #2 }}

\newcommandx{\thiswillnotshow}[2][1=]{\todo[disable,#1]{#2}}
\newcommandx{\red}[1]{{\color{red} #1}}
\newcommandx{\blue}[1]{{\color{blue} #1}}
\newcommandx{\green}[1]{{\color{green} #1}}
\newcommandx{\violet}[1]{{\color{violet} #1}}

\definecolor{lightgray}{gray}{0.8}
\definecolor{lightblue}{rgb}{0.9,0.9,1.0}
\newcommand{\grayb}[1]{\colorbox{lightgray}{#1}}
\newcommand{\blueb}[1]{\colorbox{lightblue}{#1}}

\newcommand{\eg}[0]{\textit{e.g., }}
\newcommand{\ie}[0]{\textit{i.e., }}
\newcommand{\etc}[0]{\textit{etc. }}
\newcommand{\wrt}[0]{\textit{w.r.t. }}

\usepackage{authblk}
\author[1]{Gnatyshak, Dmitry \thanks{dmitry.gnatyshak@bsc.es}}
\author[2]{Garcia-Gasulla, Dario \thanks{dario.garcia@upc.edu}}
\author[1]{\'{A}lvarez-Napagao, Sergio \thanks{sergio.alvarez@bsc.es}}
\author[2]{Arjona, Jamie \thanks{jamie.arjona@upc.edu}}
\author[3]{Venturini, Tommaso}
\affil[1]{Barcelona Supercomputing Center (BSC)}
\affil[2]{Universitat Politècnica de Catallunya (UPC)}
\affil[3]{Centre for Internet and Society CNR}

\begin{document}

\date{}
\title{Healthy Twitter discussions? Time will tell}
\maketitle

\begin{abstract}
Studying misinformation and how to deal with unhealthy behaviours within online discussions has recently become an important field of research within social studies.
With the rapid development of social media, and the increasing amount of available information and sources, rigorous manual analysis of such discourses has become unfeasible. Many approaches tackle the issue by studying the semantic and syntactic properties of discussions following a supervised approach, for example using natural language processing on a dataset labeled for abusive, fake or bot-generated content. Solutions based on the existence of a ground truth are limited to those domains which may have ground truth. However, within the context of misinformation, it may be difficult or even impossible to assign labels to instances. In this context, we consider the use of temporal dynamic patterns as an indicator of discussion health. Working in a domain for which ground truth was unavailable at the time (early COVID-19 pandemic discussions) we explore the characterization of discussions based on the the volume and time of contributions. First we explore the types of discussions in an unsupervised manner, and then characterize these types using the concept of ephemerality, which we formalize. In the end, we discuss the potential use of our ephemerality definition for labeling online discourses based on how desirable, healthy and constructive they are.
\end{abstract}


\dmitry{Tommaso needs to fill his data as an author}

\section{Introduction}

As the volume of online content and discussions grows, the amount of misinformation grows with it. The most extreme type of misinformation (content created with malicious intent), which includes fabricated or manipulated data, can be automatically identified in certain domains (\eg bot detection, image deep fake analysis), and is target of extensive research. On the other hand, less explicit types of misinformation (harmful content not necessarily produced for that purpose), such as misleading, biased or incomplete content, are much harder to characterise or label, and its characterization remains as an open challenge. Moreover, this sort of misinformation is hard to quantify, as ground truth labels are typically unavailable.

\begin{wrapfigure}{R}{0.35\textwidth}
        \vspace{-10pt}
        \begin{center}
        \includegraphics[width=0.25\textwidth]{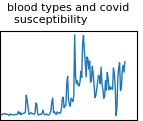}
        \end{center}
         \vspace{-10pt}
        \caption{Example of daily number of tweets mentioning the different blood types in relation with COVID-19.}
        \label{fig:example_freq}
\end{wrapfigure}

The SARS-CoV-2 pandemic illustrated how complicated it can be to identify informative debates in a context of uncertainty, and how necessary a tool is for aiding consumers at managing the overflow of information they are subject to. This context raises the following questions: What discussions can be considered healthy? Do healthy discussions have any specific patterns or properties? Is it possible to define some formal criteria for healthiness of online discussions?

For answering these questions, we collected a significant amount of tweets from the early COVID-19 pandemic period. Within this collection, we gather tweets belonging to a given thematic conversation (a topic) by defining a set of keywords specific to such theme. We characterize these \textit{topics} by looking solely at the volume and time of their activity, as shown in Figure \ref{fig:example_freq}. Given these topic representations, we cluster them based on their similarity, trying to identify main characteristics that make some topics different from others. Next, under the hypothesis that the ephemerality of conversations is related with their quality, we formalize different measures of ephemerality, and then compute their values for the gathered topics. Lastly, we explore the relation between the ephemerality measures and the unsupervised clusters. Our results indicate that ephemeral topics represent a big family of short-lived and burst-like discussions (unlikely to be informative), while non-ephemeral topics correspond to sustained, persistent and argued discussions that last on time (with potential to be informative and healthy).

\section{Related work}\label{sec:sota}

Analyzing and increasing the quality of discussions has been a topic of research since the popularization of  online platforms. At the most basic, analysis may come down to simply aggregating the events of interest and counting relevant intrinsic measures (like the number of views or reads).
For instance, \textit{altmetrics} approach for the analysis of coverage of scientific publications on social media involves collecting and calculating different metrics across various platforms \cite{altmetrics}.
The metrics used in these case come directly from the studied platforms, \ie the number of tweets and retweets for Twitter, number of comments for Facebook, \etc

One possible way for a more in-depth analysis of online discussions is analyzing their participants.
There is a number of works in this area, but it appears that although some global dynamics of the discussions can have discernible patterns, the behaviour of individual users is more or less arbitrary. Moreover, a number of bot-detection techniques rely on this property, as bots on contrary have detectable regularities in their behavior \cite{Mazza2019, chavoshi2016debot, cresci2017social, liu2017holoscope}.

Restricting the scope to news articles distribution and discussion also allows to apply more sophisticate analysis.
One of the most important characteristics of online content studied in the literature is how much attention it attracts and what is its temporal dynamics.
As it appears, the amount of attention an online user can distribute between different pieces of content is a limited resource and content creators compete for it, showing specific dynamics \cite{venturini2019, castaldo2020junk}.
A number of studies has focused on defining the types of news and laws used to generate attention (for instance, baiting high initial response, or building up their audiences and providing in-depth insights on relevant topics) \cite{Castillo2014, leskovec2009, Crane15649, Wu2007, lorenz2019accelerating, bandari2012pulse}.

Different temporal dynamics of discussion topics bring up the question of whether we can say something about the content, its types and quality (\eg the aforementioned healthiness) based purely on its temporal dynamics \cite{yang2011}.
The first stage in this would be finding ways to mathematically compare different topics based purely on the shape of their attention curves (\ie based on the dynamics of their number of tweets, views, reads, \etc).
Afterwards a vast array of methods may be used to analyze topics' similarities and differences, like time series analysis or clustering techniques.
One of the main focuses of this paper is the latter: comparing temporal distributions of various discussion topics and trying different clustering approaches to find common patterns and groups among them.

Moreover, instead of trying to analyze topics as a group, we might benefit from designing some measures that estimate the characteristic of attracted interest or attention.
In this paper we will focus on the ephemerality characteristic, originally proposed for YoutTube videos \cite{venturini2019, castaldo2020junk}, that represents how long the video in question can keep viewers' attention.

\section{Data}\label{sec:data}

\begin{wrapfigure}{R}{0.6\textwidth}
  \centering
  \includegraphics[width=\linewidth]{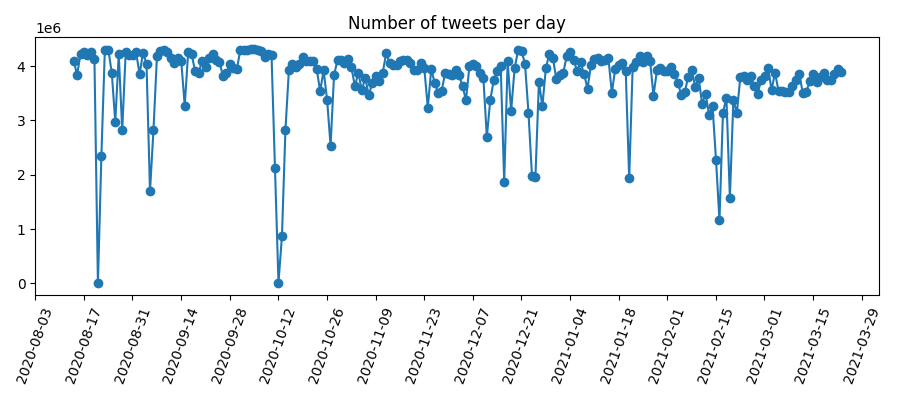}
  \caption{Gathered tweets per day in millions}
  \label{fig:mongo-tweets}
  \vspace{-10pt}
\end{wrapfigure}

The data collection process used in this work can be divided into three sequential stages. The first and second stages are designed to obtain a large, representative set of online discussions. One that includes a varied set of social behaviours in times of uncertainty. The third stage creates subsets of tweets belonging to a given topic, typically one that generates a discussion sustained through time. Let us define these stages in further detail:

\begin{enumerate}
    \item Accessing and gathering online tweets through the Twitter API, using a query made up of 30 keywords related with COVID-19 (see those in Table~\ref{tab:col-keywords}). We used a \textit{MongoDB}\footnote{\url{https://www.mongodb.com/}} database for storage and a Python script accessing the Twitter stream for the gathering. 
    
    \item Indexing tweets, enabling quick text search. This is necessary considering our dataset contains 829 million tweets, spanning seven months (between August 15, 2020 and March 24, 2021). The global distribution of gathered tweets on day-by-day basis is presented on Figure~\ref{fig:mongo-tweets}, with a mean of 3,735,678 tweets per day, and a standard deviation of 667,488. For this we used a \textit{Solr}\footnote{\url{https://lucene.apache.org/solr/}} database.
    
    \item Extracting subsets of tweets corresponding to topics, from the general collection of tweets. This is detailed in \ref{subsec:topic_gen}.
\end{enumerate}

\begin{table}[ht!]
\caption{Collection of tweet keywords used for the general query.}
\label{tab:col-keywords}
\begin{tabular}{cccccc} 
\hline
covid     &  cov-19          &  sarscov2      &  flu        &  zoom         &contagious\\
covid19   &   cov19       &  sars-cov-2      &  epidemic         &   zoommeeting                & fever\\
covid-19  & covd19       & coronavirus         &  pandemic             &   social distance           &cough\\
covid\_19 &  covd 19       & corona virus  &  outbreak        &  clapbecausewecare          &symptoms\\  
2019-ncov & 2019ncov         &  hantavirus                &  lockdown   &   infection       &  stayhome     \\
\hline
\end{tabular}
\end{table}

Besides the bodies of tweets, we also gathered all existing metadata. This includes the tweet author information and the number of times that a tweet has been retweeted, quoted or replied.

\subsection{Topic generation}\label{subsec:topic_gen}

Once raw tweets have been gathered and stored, we proceed to extract sets of tweets corresponding to specific topics. We identify COVID-19-related controversies, misinformation topics and rumours from Poynter's database\footnote{\url{https://www.poynter.org/ifcn-covid-19-misinformation/}}.
To represent each of these topics, we manually composed a \textit{Solr} query for each of them based on their description. The goal is to retrieve the relevant tweets for a topic using their most characteristic keywords. Follows an example of \textit{Solr} query, one written to extract the discussion topic around the combination of Ibuprofen and COVID-19:

\vspace{7pt}
\noindent\begin{minipage}{\linewidth}
\begin{quote}

\blueb{\textbiglparen{}\grayb{covid} \textbf{OR} \grayb{covid-19} \textbf{OR} \grayb{covid19} \textbf{OR} \grayb{covid\_19} \textbf{OR} \grayb{cov-19} \textbf{OR} \grayb{cov19} \textbf{OR}}

\blueb{\grayb{covd19} \textbf{OR} \grayb{covd-19} \textbf{OR} \grayb{(covd \textbf{AND} 19)} \textbf{OR} \grayb{2019-ncov} \textbf{OR} \grayb{2019ncov} \textbf{OR}}

\blueb{\grayb{(corona \textbf{AND} virus)} \textbf{OR} \grayb{coronavirus} \textbf{OR} \grayb{cvd-19} \textbf{OR} \grayb{sars-cov-2} \textbf{OR} \grayb{sarscov2}\textbigrparen{}}

\begin{center}
    \textbf{AND} \blueb{ibuprophen}
\end{center}

\vspace{0.1cm}
\end{quote}
\end{minipage}
\dario{I would clarify that there is a basic query for covid (and show it separately), and then explain that we add an AND clause for the topic at hand defined by a set of AND joined terms. Be sure to clarify this query is different from the one in table 1}
\dario{Add an annex (that may or may not go with this paper) with the 78 topics and their queries}

In total, we characterized 68 different misinformation topics within a seven month period. See Table~\ref{tab:topics} for a sample of topics and their statistics. We represented each of those topics as an integer vector of 222 positions, where each element contains the number of tweets for that topic in the corresponding day. We call them topic distribution vectors (TDV) \jam{to agree with dario (15/03/2021)}, and intuitively they represent the attention dynamics of society for the corresponding topics.

\begin{table}
\caption{Daily statistics (average and quartiles) for a sample of topics}
\label{tab:topics}
\begin{tabular}{@{}llrllll@{}}
\toprule
\multicolumn{1}{c}{\textbf{Title}} & 
\multicolumn{1}{c}{\textbf{\# tweets}} & 
\multicolumn{1}{c}{\textbf{Daily avg}} & 
\multicolumn{1}{c}{\textbf{Q1}} & 
\multicolumn{1}{c}{\textbf{Q2}} &
\multicolumn{1}{c}{\textbf{Q3}} & \\ \toprule
(cure or vaccine) and covid   & 39,303,087    & 177,040.93 & 92,307.0 & 173,076.5 & 239,246.25 \\ \midrule
lockdowns and covid         & 9,092,966     & 40,959.31   & 28,583.0 & 37,926.0 & 49,314.25 \\ \midrule
flu and covid           & 3,917,950     & 17,648.42   & 8,971.75 & 14,402.0 & 21,659.75 \\ \midrule
invermectin and covid   & 290,662       & 1,309.29   & 144.75 & 701.0 & 2296.25 \\ \bottomrule
\end{tabular}
\end{table}

\subsection{Pre-processing}\label{subsec:preprocessing}

According to official Twitter documentation, the \textit{real-time stream and filtered API}\footnote{\url{https://dev.to/twitterdev/stream-tweets-in-real-time-with-v2-of-the-twitter-api-46fc/}} provides data filtered by our original COVID-19 query in a continuously form (each second) with the limitation that the amount of data returned by the query cannot be greater than the 1\% of the total number of tweets in that moment. In that case, Twitter sends a notification with the number of exceeding tweets. \jam{a revisar per en Dario. Added documentation as footnote and explained how it works.} While this amount is still significant in volume (see Figure~\ref{fig:mongo-tweets}), this may entail a certain amount of arbitrary variance in our data. Furthermore, since we use \textit{one day} as our atomic unit of measure, holidays, weekends and other special dates may add noise to the TDV \jam{dario...}. 

\begin{figure}
  \centering
  \includegraphics[width=0.98\textwidth]{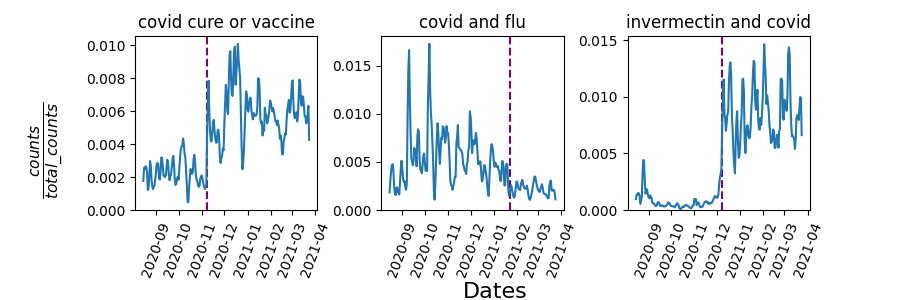}
  \caption{Normalized and smoothed frequency curves for a sample of topics.}
  \label{fig:preprocess-topics}
\end{figure}

To mitigate the impact of variance and noise, while minimizing the loss of precision, we apply a smoothing technique on vectors. For that we use a three-day sliding window average, reducing this way the impact of weekly and weekend patterns. To equal the volume of all vectors (\eg some discussions may last longer and engage a larger amount of contributions) we normalize them so that the content of each vector sums up to one. 
Figure \ref{fig:preprocess-topics} shows some TDV after applying smoothing and normalization. The doted purple line indicates the date with a significant change of the volume of tweets for the related topic. In the case of the \textit{(cure or vaccine) and covid} topic, the volume change on November coincides with Pfizer announcement of the vaccine\footnote{\url{https://www.businesswire.com/news/home/20201109005539/en/}}. In the case of the \textit{flu and covid} topic it is clear that there is a seasonal influence on the volume of tweets as the flu is more common on winter. Lastly, the \textit{invermectin and covid} topic shows a change in volume in December, it can be related to a video clip posted online were a medical doctor described ivermectin as a “wonder drug”\footnote{\url{https://apnews.com/article/fact-checking-afs:Content:9768999400}}.
\section{Distances and Clusters of Topic Distribution Vectors}\label{sec:basic-analysis}

According to \cite{venturini2019}, the evolution of discussion volume over time is related with the reliability, quality and constructiveness of online debates. We explore this hypothesis by building a set of TDV which represent specific debates during the COVID-19 pandemic. Since we work in a domain without a reliable ground truth, and we wish to apply our findings to any dataset (regardless of size or expert review), we focus on an unsupervised approach with the goal to understand how do temporal dynamics characterize the healthiness of topics.

The first step in the process is to explore the notion of distance between pairs of TDV. In \ref{subsec:similarity} we consider different metrics, which illustrates the need for a topic alignment policy. This is discussed in \ref{subsec:alignment}. Finally, we conduct a series of clustering experiments in \ref{subsec:clustering}, with which we seek to identify set of discussions with distinct temporal behaviors.

\subsection{Topic Similarity}\label{subsec:similarity}

As mentioned in \S\ref{sec:data}, the nature of the data (number of tweets related to a topic for each day within a period of 222 days) makes that the minimum value for each element of a TDV is 0 and, because of the usage of normalization, the sum of its elements sum up to 1. Such representation of the data is equivalent to the definition of probability distribution function (i.e: the minimal probability value for an outcome is 0, all values are between 0 and 1, and the sum of the probabilities of all the outcomes sum up to 1). Considering these similarities, we have adapted distance measures typically used for probability distributions in order to be used to compare the similarities between the temporal behaviour dynamics of the TDV.
\jam{dario How you see now the justification of the usage of such metrics?}

\textbf{Sum of absolute differences (SAD).}
A straight-forward method to define distances between topic vectors is to calculate the sum of absolute element-wise differences. Its lower bound is naturally 0 (\ie two topics which perfectly overlap), and if we normalize it by 0.5 coefficient, its upper bound will be 1 as the maximal difference between 2 vectors each of which sums up to 1 is 2 (\ie two topics with no overlap).
Let $t_i$ and $t_j$ be two topic vectors of length $M$.
Let $t_{im}$ denote $m$\textsuperscript{th} element of vector $t_i$.
Than we can define SAD distance as:

\begin{equation}
    d_{SAD}\left(t_i, t_j\right) = \frac{1}{2} \sum_{m=1}^{N}{\left|t_{im} - t_{jm}\right|}
\end{equation}

The main drawback of this distance metric is its strong connection to exact dates. Even if two topics have their tweet number curves identical, the distance may be large if they are shifted temporally. This happens naturally in our dataset, as discussion topics start or end on independent dates. Thus, before SAD can be used, we need to align the topic vectors, as discussed in \S\ref{subsec:alignment}. Furthermore, this metric is very sensitive to noise and perturbations in the shapes of topic distributions (\eg punctual outliers carry a lot of weight). For this reason, we consider other distances which are more resilient to spurious variations.

\textbf{Kolmogorov-Smirnov statistic adaptation (KS).} 
In its original form, the Kolmogorov-Smirnov (KS) statistic is calculated for two empirical distribution functions (EDF) (or on an EDF and cumulative distribution function (CDF)) to determine whether the underlying distributions are the same. Essentially, KS statistics equals to the maximal absolute difference between the components of two EDFs, which is bound to $[0, 1]$. Since the normalized topic vectors closely resemble empirical probabilities, we may perform the same calculation on them. In order to get an adapted KS distance we need to turn the topic vectors $t_i$ and $t_j$ into cumulative vectors $\hat t_i$ and $\hat t_j$. Then, KS is calculated as follows:

\begin{equation}
    d_{KS}\left(\hat t_i, \hat t_j\right) = \max_{1\leq m\leq N}{\left| \hat t_{im} - \hat t_{jm} \right|}
\end{equation}

KS is less sensitive to noise and small perturbations than SAD because it focuses on the differences between maximal accumulated tweet masses for the topics. Like SAD, KS suffers from the temporal misalignment of topics.

\textbf{Hellinger distance adaptation (HDA).}
The Hellinger distance is a metric used to evaluate the difference between two probability distributions by computing the difference between square roots of probabilities:

\begin{equation}
    d_{HDA}\left(t_i, t_j\right) = \frac{1}{\sqrt{2}} \sqrt{\sum_{m=1}^{N}\left(\sqrt{t_{im}} - \sqrt{t_{jm}}\right)^2}
\end{equation}

This is a metric bound within $[0,1]$ interval were 0 means that the probability distributions are equal and 1 that they are totally different. In this case, 1 is achieved when for every positive value of the probabilities of one of the distribution the other has 0 probability. One problem of this metric is that the use of square roots makes it sensitive to values close to 0, making the days with less activity most relevant for the distance.

\textbf{Norm of difference of squares (NDS).}
Just as the Hellinger distance increases the importance of the low-value elements of the topic vectors by using square roots, we consider the opposite approach: lower the significance of lower values by using powers. This way the values of the vectors' elements get pushed towards 0, and the lower the values~--- the more they get decreased. In this case the metric will focus more on areas with high peaks and less on areas with low perturbations.

\begin{equation}
    d_{NDS}\left( t_i, t_j\right) = \frac{1}{\sqrt{2}} \sqrt{\sum_{m=1}^{N}\left(t_{im}^2 - t_{jm}^2\right)^2}
\end{equation}

This metric is bounded to the $[0,1]$ interval and, like the previous metrics, is dependent on the vectors alignment but it serves well to our purposes because lower values on the TDV can be related to noise and are not adding valuable information about the characteristics of the discussion.

\subsection{Topic alignment}\label{subsec:alignment}

As have been observed, most of the proposed distance measures require the topics to be aligned temporally. Otherwise, the distance between vectors will be mostly derived from their relative time of occurrence, which is unrelated with our goal (\ie characterizing the healthiness of discussions, regardless on when they happen in time). To overcome this issue, next we propose different methods to align the TDV.

\textbf{Highest peak (\textbf{max}-alignment).}
One of the simplest ways to align topics is to shift them with zero-padding so that the maximal element of each topic (\ie the day with most activity) corresponds to the same index. The main issue with max-alignment is its arbitrary nature. Since we can make no assumptions over data distribution, the highest peak can easily shift from one extreme of the temporal range to the other (consider a \textit{'U'} shaped topic distribution, and how it would drastically affect the metric to align either by the first or the last peak).

\begin{wrapfigure}{R}{0.65\textwidth}
  \centering
  \includegraphics[width=\linewidth]{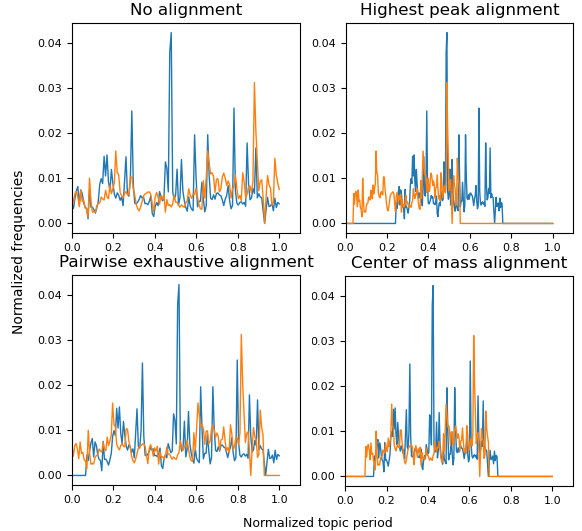}
  \caption{Example of different alignments for two topics.}
  \label{fig:alignments}
\end{wrapfigure}

\textbf{Center of mass (\textbf{mean}-alignment).}
Another approach used to align different probability mass functions is to compute the center of mass of each of the TDV and make the alignment on those. First, the total mass of each TDV is computed as the sum of each element of the topic distribution vector (frequency for a date) multiplied by its position in the vector. Then it is divided by the length of the topic distribution vector of each topic vector calculated as the mean of dates weighted by their tweet frequency values. This is a more robust solution than the highest peak, since the center of mass may not change as dramatically by adding or subtracting a few data samples from the topic. Nonetheless, misalignment is still possible as the same center of mass can correspond to two vectors with very different shapes, for instance, when one TDV has its center of mass defined by one burst of tweets at the middle of the length of the vector and another one has two burst of tweets at equal distances from that point the result will be the same center of mass but the vectors will still be misaligned.
\dario{dont quite get the example. can it be improved?} \jam{reworked, can you check it dario?}.

\textbf{Pairwise exhaustive alignment.}
Finally, it is possible, albeit computationally demanding, to find the alignment that minimizes the distance between each pair of topics. This requires to exhaustively compute the distance between a couple of vectors for every possible alignment setting. Considering the simplicity of our data (\ie 222-long vectors) computing a pairwise exhaustive alignment is feasible. Notice this solution may potentially use different alignments for each distinct pair of vectors.

An example of the different alignments are shown in Figure~\ref{fig:alignments} for two topics where its burst happens at different days.

\dario{a couple of final sentences mentioning which and how will these alignments be used}

Due to limitations of the max-alignment and the mean-alignment and the fact that pairwise exhaustive alignment with vectors of length 222 is computational feasible, we have chosen the latter in order to proceed with the analysis. This means that the pairwise exhaustive alignment is used with the NDS distance. The resulting minimal distances are then used as to perform the clustering technique with the goal of obtain clusters that can be representative of healthy and unhealthy conversations.
\subsection{Clustering}\label{subsec:clustering}

In order to characterise the different topics based on their temporal dynamics, and without any assumption regarding their validity, we perform clustering using the computed distances. As mentioned, we used the NDS distance on the vectors aligned exhaustively, after normalization and smoothing these using a three day sliding window.

\begin{figure}[tb!]
    \centering
    \begin{subfigure}[b]{.45\linewidth}
        \centering
        \includegraphics[width=\linewidth]{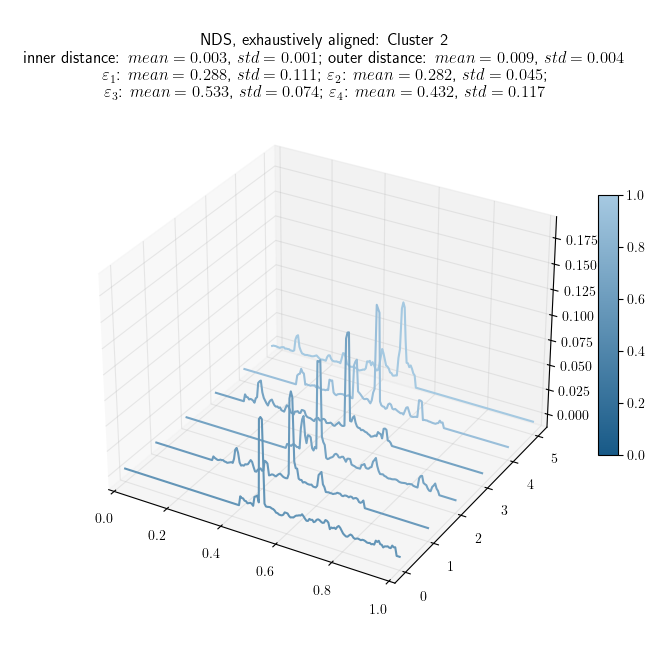}
        \caption{Burst topics cluster}
    \end{subfigure}
    \begin{subfigure}[b]{.45\linewidth}
        \centering
        \includegraphics[width=\linewidth]{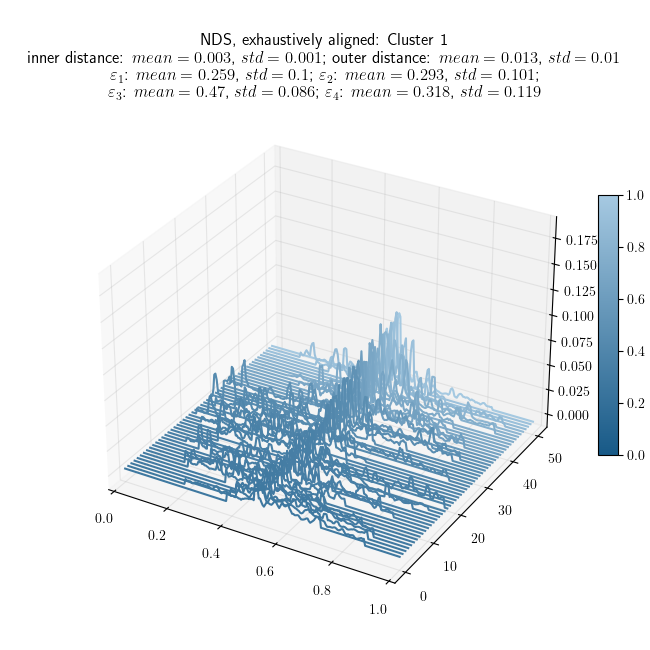}
        \caption{``Uniform'' topics cluster}
    \end{subfigure}
    \caption{Clusters with full data preprocessing, exhaustively aligned, and using NDS distance}
    \label{fig:p2fullpreclus}
\end{figure}

The chosen clustering method is HDBSCAN, a density and hierarchy method which uses a precomputed distance matrix to find the best number of clusters. This algorithm was chosen because it permits to use a distance matrix instead of a dataset and provides the optimal number of clusters with the only requirement, apart from the distance matrix, of choosing a minimal number of elements in each cluster (which we set up to \jam{justify?}3). HDBSCAN internally computes the minimum spanning tree and converts it into the hierarchy of connected components, sorting the distances of the tree in increasing order from individuals to a single cluster. With the obtained hierarchy and using the minimal number of elements for forming a cluster the hierarchy tree becomes much smaller. The last step is to extract the number of clusters, HDBSCAN defines internally the stability of each of the clusters as the summation of the time that points belongs to the cluster, this is computed as the difference of the inverse of the distance in which a point stops being part of a cluster with the inverse of the distance in which a point started to belong to that cluster. If the stability of a cluster is greater than their children clusters, then it is a candidate to be a selected cluster. Once arriving at the root, we get the clusters of our distance matrix as those with maximum stability.\jam{FW add dendograms to support this explanation about HDBSCAN}

Results show three communities: two with distinct visual features and sizes (51 and 6 topics) and one with outliers topic vectors. The distinct clusters correspond to either topics with one large peak or burst of activity (the small one), and topics that are generally more uniform in its distribution of activity (the large one). The outliers cluster includes everything that does not fit these two definitions.

The clusters obtained by HDBSCAN are shown in Figure~\ref{fig:p2fullpreclus} along with some additional statistics. The color of each topic denotes the normalized  distance from it to the cluster centroid (with 0 being the cluster center and distance 1 being the most distant cluster member from it).\dario{FW:change color gradient. not visible as it is}\dario{are topics sorted in depth by distance to centroid? it looks like it}\dario{FW increase vertical scale of plots to increase visibility} Visually, burst topics seem to be characterized by a large, central peak of activity, surrounded by an area of diminished contributions. On the other hand, uniform topics show a more balanced distribution. Notice how, within the uniform cluster, those topics that have a bigger peak are those further from the uniform cluster centroid (\ie those in a clearer color, deeper into the figure). These borderline cases further reinforces the intuitive difference between burst and uniform clusters.

\section{Ephemerality}\label{sec:ephemerality}
\jam{15/03/2022 proof reading with love on Monday. In the conclusions ephemerity 3 is mentioned, but I think that we need to consolidate (reindex) the ephemerities and redo some figures (or gimp them..). To be commented with @dario}
After having identified two distinct clusters of online discussions based on the topic distribution vectors, we visually hypothesize that these clusters are composed by either burst-like topics on one hand, and more uniform topics on the other. To reinforce this notion, while exploring the definition of the term, we now consider the \textit{ephemerality} of topics and its relation with the found clusters.

Among the works exploring online discussions, a few have focused on the temporal aspects of it\cite{}\jam{do not forget}, \ie looking at how discussion evolve over time in volume and quality\jam{qualiy or quantity?}. In this context, a concept of particular relevance is \textit{ephemerality}~ \cite{venturini2019, castaldo2020junk}. Originally proposed for the analysis of attention in online videos, ephemerality describes how much time it takes to accumulate most (empirically set to 80\%\jam{15-03-2022 motivate this}) attention or interactions when considering the total attention gathered. 

Considering the volume of our database (counting millions of tweets per day), ephemerality analysis, which relies on voluminous data, is an appropriate approach. In our context, ephemerality is based on the number of days passed between the first appearance of tweets for a topic, to the day when that topic accumulates 80\%\jam{15-03-2022 motivate this} of all tweets. This interpretation of ephemerality measures the temporal distribution of tweets generated. Since there are no formal definitions of ephemerality for our context, we first propose some alternatives. 

\subsection{Formalization}

Let us now formally define our ephemerality measures. Let $t_i$ denote a normalized frequency vector of topic $i$ of length $M$; let $t_{im}$ denote its $m$\textsuperscript{th} element. According to the original definition of ephemerality~\cite{}, for our data it would formalize as follows:

\begin{equation}
    \varepsilon_{orig}\left(t_i\right) = 1 - \frac{\left(\min{m}: {\sum_1^m{t_{im}} \ge 0.8}\right) - \left(\min{m}: {t_{im} > 0}\right) + 1}{\left(\max{m}: {t_{im} > 0}\right) - \left(\min{m}: {t_{im} > 0}\right) + 1}
\end{equation}

Here we compute the proportion of time taken by a topic to reach 80\% of activity, with respect to the topic period of activity. We then subtract the result from 1 to compute ephemerality (closer to 1 more ephemeral). This definition allows us to compare the ephemeralities of topics of different duration. The main drawback of this approach is its sensitivity to outliers. A single early tweet matching the topic query keywords will specify the starting date of the topic, and heavily influence the proportion of time taken until reaching 80\% of activity. Furthermore, this approach may provide different ephemerality scores to two burst topics, if their corresponding bursts happen either close to the beginning or close to the ending of the topic lifespan.

One way to limit the impact of both factors is to filter out data at both ends of the topic distribution vector. For example by removing 10\% of tweets from either side, and then analyzing the (relative) length of the ``middle'' section of the discussion.

\begin{equation}
    \varepsilon_{2}\left(t_i\right) = 1 - \frac{\left(\min{m}: {\sum_1^m{t_{im}} \ge 0.9}\right) - \left(\min{m}: {\sum_1^m{t_{im}} \ge 0.1}\right) + 1}{\left(\max{m}: {t_{im} > 0}\right) - \left(\min{m}: {t_{im} > 0}\right) + 1}
\end{equation}

According to this definition, ephemerality values are limited between 0.2 (least ephemeral) and 1 (most ephemeral). Zero ephemerality is unreachable since we are discarding a 20\% of activity on both ends, making $varepsilon_{2}=1-\frac{0.8}{1}$ the minimum. While this solution is more resistant to the relative position of bursts (\ie burst topics with the peak near the beginning or the end with score similarly), and to the arbitrary occurrence of the first and last activity, it may be affected by multiple bursts (\ie having an early and a late burst will result in low ephemerality). To fix that, we consider a definition of ephemerality which gets rid of the temporal dimension, sorting the topic frequency vector in descending order of activity, and calculating the ephemerality on that sorted vector:

\begin{equation}
    \varepsilon_{4}\left(t_i\right) = 1 - \frac1{0.8} \cdot \frac{\min{m}: {\sum_1^m{\widehat{t}_{im}} \ge 0.8}}{\max{m}: t_{im} > 0}
\end{equation}
\dario{rename ephemeralities in figures and text. 2=filtered 4=sorted?}
where $\widehat{t}$ denotes the sorted array of tweet frequencies for the topic $m$.
Using this form of ephemerality breaks the connection of the frequencies to the specific days and allows us to see if the discussion contained enough of days with enough tweets. The limitation of this approach is that it considers only the proportion of days with most activity, regardless of their relative position (\ie ephemerality is the same for topic where all activity happens in four burst days, regardless of these days being close or far from one another in time).

\begin{wrapfigure}{R}{0.4\textwidth}
  \includegraphics[width=0.38\textwidth]{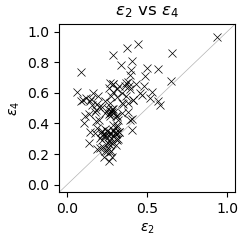}
  \caption{Ephemerality values plotted against each other}
  \label{fig:ephemeralities-corr}
  \vspace{-20pt}
\end{wrapfigure}

\begin{figure}[bh!]
  \centering
  \includegraphics[width=.95\textwidth]{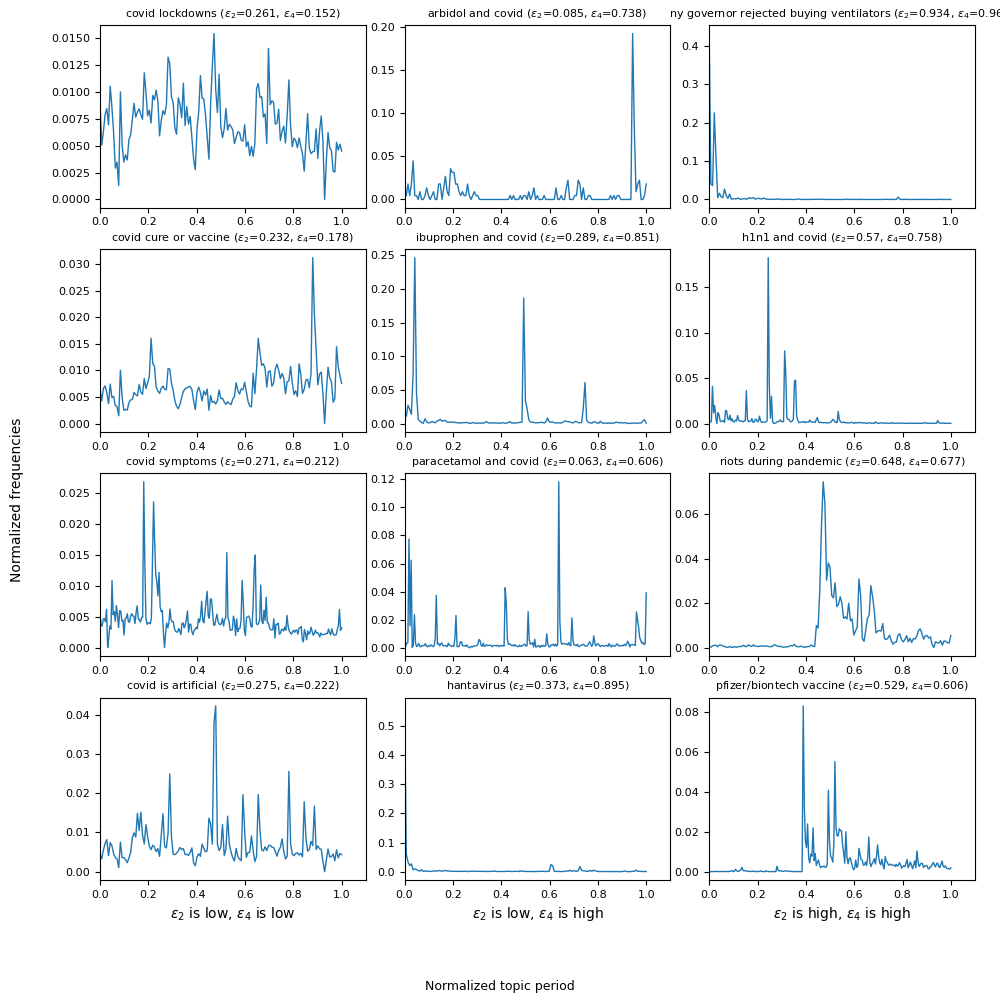}
  \caption{Examples of topics with different combinations of $\varepsilon_2$ and $\varepsilon_4$ values. Note that due to the topic selection, the last column contains topics with comparatively low $\varepsilon$ values.}
  \label{fig:e-cross-examples}
\end{figure}

Considering the properties and limitations of $varepsilon_{2}$, and $varepsilon_{4}$, and how they complement each other, we decide to use both. A correlation analysis shows how both metrics provide distinct information (see Figure~\ref{fig:ephemeralities-corr}). While $varepsilon_{2}$ focuses on the length of the central part of the distribution, it is insensitive to the shape of that middle section. On the other hand, $varepsilon_{4}$ counts in how many days the topic concentrated its activity, but it does so at cost of properly detecting temporal length of the discussion. Table~\ref{tab:ephemeralities} shows how we can interpret the combinations of ephemerality values, and Figure~\ref{fig:e-cross-examples} shows samples of topics falling within these categories. Here by \textit{burst} topics we mean the topics with a single high peak lasting for one or several days, but comparatively short \wrt the topic duration, and by \textit{rollercoaster} topic~--- one with at least two burst separated by periods of low activity.


\begin{table}[tb]
\caption{The shape of topic's frequency vector based on the ephemerality values}
\begin{tabular}{|l|l|l|}
\hline
 & \bm{$\varepsilon_{4}$} is \textbf{low} & \bm{$\varepsilon_{4}$} is \textbf{high} \\ \hline
  \bm{$\varepsilon_{2}$} is \textbf{low} & Uniform and sustained topic & Rollercoaster topic \\ \hline
\bm{$\varepsilon_{2}$} is \textbf{high} & --- & Single burst/peak topic \\ \hline
\end{tabular}
  \label{tab:ephemeralities}
\end{table}









\subsection{Ephemerality and cluster characterization}

We explore the relation between the ephemeralities defined in this section, and the clusters found in \ref{subsec:clustering}. Statistics can be seen in Figure~\ref{fig:p2fullpreclus}.

Cluster 1 includes topics with lower $\varepsilon_4$ values (mean 0.318, std 0.119) than Cluster 1 topics (mean 0.432, std 0.117). Meanwhile, $\varepsilon_2$ values seem to be rather similar between both clusters (mean 0.293 vs mean 0.282), although Cluster 1 includes a higher variance (std 0.101 vs std 0.045). According to these statistics, Cluster 2 contains most burst-like topics, and Cluster 1 most uniform topics. However, the difference between single-burst and rollercoaster topics is not properly represented with these two clusters.

\section{Conclusions and future work}\label{sec:conclusion}

In this paper we aimed at analyzing online discussions on various topics using only non-semantical temporal information on user attention to them.
Our objective was to figure out which discussions can be considered healthy, whether it is possible to extract this information from the aforementioned attention distributions, and, if so, what would be formal measures to estimate it.

The work resulted in two proposals.
First of all, our experiments show that for the purpose of clustering using HDBSCAN method, the best quality \jam{15-03-2021 but we have decided based on limitations of each proposal...} of the produced clusters is achieved with the proposed NDS distance measure applied for pairwise aligned distribution vectors.
Normalization is needed so the topic distribution vectors can be compared using metrics proposed to compare distribution. Smoothing is optional, although one has to be careful, as it might decrease the magnitude of peaks which in turn would move some bursty topics into ``uniform'' clusters. The NDS distance measure has been chosen because it is insensitive to low frequency values that can be related to noise. Because an alignment is needed, from the three proposed alignments, the pairwise exhaustive alignment is the one chosen as it provides the minimal distance between topics based on the chosen metric.

Secondly, we have formalized the notion of ephemerality, that estimates the ability of a discussion topic to maintain users' attention.
We have studied different types of ephemerality measures, aiming at different aspects of what topics might be considered ephemeral; $\varepsilon_3$\jam{OJO!} showed the best results as a standalone measure with a combination of $\varepsilon_2$ and $\varepsilon_4$\jam{subindex to be fixed} (the former focuses on topics that last longer, the latter~--- on topics with higher number of days with high attention) being the close second.

Although the current results are promising, there is a number of future research lines:

\begin{itemize}
    \item First of all, it may be beneficial to ensure that the discussion chains of tweets are kept as intact as possible (which will require additional parsing of Twitter data and extension of the exisitng topics).
    \item Secondly, more clustering methods can be tested with this use-case.
    \item Thirdly, the ephemerality threshold must not be set specifically to 80\%. It might prove useful to try other values or implement some form of a dynamic threshold.
\end{itemize}

\dmitry{FW: try to calculate and subtract the noise from topics? Or at least consider all the absolute frequencies less than a certain number as 0}

\section*{Acknowledgement}
  This work is supported by the scheme `INFRAIA-01-2018-2019: Research and Innovation action', Grant Agreement n. 871042 `SoBigData++: European Integrated Infrastructure for Social Mining and Big Data Analytics'.

\printbibliography 

@misc{castaldo2020junk,
      title={Junk News Bubbles: Modelling the Rise and Fall of Attention in Online Arenas}, 
      author={Maria Castaldo and Tommaso Venturini and Paolo Frasca and Floriana Gargiulo},
      year={2020},
      eprint={2004.08863},
      archivePrefix={arXiv},
      primaryClass={cs.SI}
}

@incollection{venturini2019,
  TITLE = {{From Fake to Junk News, the Data Politics of Online Virality}},
  AUTHOR = {Venturini, Tommaso},
  URL = {https://hal.archives-ouvertes.fr/hal-02003893},
  BOOKTITLE = {{Data Politics: Worlds, Subjects, Rights}},
  EDITOR = {Didier Bigo and Engin Isin and Evelyn Ruppert},
  PUBLISHER = {{Routledge}},
  YEAR = {2019},
  HAL_ID = {hal-02003893},
  HAL_VERSION = {v1},
}

@article{altmetrics,
    author = {Zahedi, Zohreh AND Costas, Rodrigo},
    journal = {PLOS ONE},
    publisher = {Public Library of Science},
    title = {General discussion of data quality challenges in social media metrics: Extensive comparison of four major altmetric data aggregators},
    year = {2018},
    month = {05},
    volume = {13},
    url = {https://doi.org/10.1371/journal.pone.0197326},
    pages = {1-27},
    number = {5},
    doi = {10.1371/journal.pone.0197326}
}

@inproceedings{Castillo2014,
author = {Castillo, Carlos and El-Haddad, Mohammed and Pfeffer, J\"{u}rgen and Stempeck, Matt},
title = {Characterizing the Life Cycle of Online News Stories Using Social Media Reactions},
year = {2014},
isbn = {9781450325400},
publisher = {Association for Computing Machinery},
address = {New York, NY, USA},
url = {https://doi-org.recursos.biblioteca.upc.edu/10.1145/2531602.2531623},
doi = {10.1145/2531602.2531623},
booktitle = {Proceedings of the 17th ACM Conference on Computer Supported Cooperative Work \&; Social Computing},
pages = {211–223},
numpages = {13},
location = {Baltimore, Maryland, USA},
series = {CSCW '14}
}

@inproceedings{leskovec2009,
author = {Leskovec, Jure and Backstrom, Lars and Kleinberg, Jon},
title = {Meme-Tracking and the Dynamics of the News Cycle},
year = {2009},
isbn = {9781605584959},
publisher = {Association for Computing Machinery},
address = {New York, NY, USA},
url = {https://doi-org.recursos.biblioteca.upc.edu/10.1145/1557019.1557077},
doi = {10.1145/1557019.1557077},
booktitle = {Proceedings of the 15th ACM SIGKDD International Conference on Knowledge Discovery and Data Mining},
pages = {497–506},
numpages = {10},
location = {Paris, France},
series = {KDD '09}
}

@article{Crane15649,
archivePrefix = {arXiv},
arxivId = {0803.2189},
author = {Crane, Riley and Sornette, Didier},
doi = {10.1073/pnas.0803685105},
eprint = {0803.2189},
issn = {00278424},
journal = {Proceedings of the National Academy of Sciences of the United States of America},
number = {41},
pages = {15649--15653},
pmid = {18824681},
publisher = {National Academy of Sciences},
title = {{Robust dynamic classes revealed by measuring the response function of a social system}},
url = {https://www.pnas.org/content/105/41/15649},
volume = {105},
year = {2008}
}

@article{Wu2007,
archivePrefix = {arXiv},
arxivId = {0704.1158},
author = {Wu, Fang and Huberman, Bernardo A.},
doi = {10.1073/pnas.0704916104},
eprint = {0704.1158},
issn = {00278424},
journal = {Proceedings of the National Academy of Sciences of the United States of America},
month = {11},
number = {45},
pages = {17599--17601},
pmid = {17962416},
publisher = {National Academy of Sciences},
title = {{Novelty and collective attention}},
url = {http://www.ncbi.nlm.nih.gov/pubmed/17962416 http://www.pubmedcentral.nih.gov/articlerender.fcgi?artid=PMC2077036},
volume = {104},
year = {2007}
}

@inbook{Mazza2019,
author = {Mazza, Michele and Cresci, Stefano and Avvenuti, Marco and Quattrociocchi, Walter and Tesconi, Maurizio},
title = {RTbust: Exploiting Temporal Patterns for Botnet Detection on Twitter},
year = {2019},
isbn = {9781450362023},
publisher = {Association for Computing Machinery},
address = {New York, NY, USA},
url = {https://doi-org.recursos.biblioteca.upc.edu/10.1145/3292522.3326015},
booktitle = {Proceedings of the 10th ACM Conference on Web Science},
pages = {183–192},
numpages = {10}
}

@inproceedings{chavoshi2016debot,
  title={Debot: Twitter bot detection via warped correlation.},
  author={Chavoshi, Nikan and Hamooni, Hossein and Mueen, Abdullah},
  booktitle={Icdm},
  pages={817--822},
  year={2016}
}

@article{cresci2017social,
  title={Social fingerprinting: detection of spambot groups through DNA-inspired behavioral modeling},
  author={Cresci, Stefano and Di Pietro, Roberto and Petrocchi, Marinella and Spognardi, Angelo and Tesconi, Maurizio},
  journal={IEEE Transactions on Dependable and Secure Computing},
  volume={15},
  number={4},
  pages={561--576},
  year={2017},
  publisher={IEEE}
}

@inproceedings{liu2017holoscope,
  title={Holoscope: Topology-and-spike aware fraud detection},
  author={Liu, Shenghua and Hooi, Bryan and Faloutsos, Christos},
  booktitle={Proceedings of the 2017 ACM on Conference on Information and Knowledge Management},
  pages={1539--1548},
  year={2017}
}

@article{lorenz2019accelerating,
  title={Accelerating dynamics of collective attention},
  author={Lorenz-Spreen, Philipp and M{\o}nsted, Bjarke M{\o}rch and H{\"o}vel, Philipp and Lehmann, Sune},
  journal={Nature communications},
  volume={10},
  number={1},
  pages={1--9},
  year={2019},
  publisher={Nature Publishing Group}
}

@inproceedings{bandari2012pulse,
  title={The pulse of news in social media: Forecasting popularity},
  author={Bandari, Roja and Asur, Sitaram and Huberman, Bernardo},
  booktitle={Proceedings of the International AAAI Conference on Web and Social Media},
  volume={6},
  number={1},
  year={2012}
}

@inproceedings{yang2011,
author = {Yang, Jaewon and Leskovec, Jure},
title = {Patterns of Temporal Variation in Online Media},
year = {2011},
isbn = {9781450304931},
publisher = {Association for Computing Machinery},
address = {New York, NY, USA},
url = {https://doi-org.recursos.biblioteca.upc.edu/10.1145/1935826.1935863},
doi = {10.1145/1935826.1935863},
booktitle = {Proceedings of the Fourth ACM International Conference on Web Search and Data Mining},
pages = {177–186},
numpages = {10},
location = {Hong Kong, China},
series = {WSDM '11}
}
\end{document}